\documentclass[sigconf]{acmart}

\usepackage{mathtools,bm}
\usepackage{booktabs}
\usepackage{placeins}
\usepackage{enumitem}
\usepackage{graphicx}
\usepackage{tabularx}
\usepackage{array}
\usepackage{microtype}

\usepackage{multirow}

\setlist[enumerate]{
  leftmargin=*,
  topsep=2pt,
  itemsep=0pt,
  parsep=0pt,
  partopsep=0pt
}

\newcommand{\RV}{\mathrm{RV}}
\newcommand{\BV}{\mathrm{BV}}
\newcommand{\GK}{\mathrm{GK}}
\newcommand{\IV}{\mathrm{IV}}
\newtheorem{proposition}{Proposition}

\acmConference[ICAIF '26]
{ACM International Conference on AI in Finance}
{November 14--17, 2026}
{Milan, Italy}
\acmBooktitle{ACM International Conference on AI in Finance
(ICAIF '26), November 14--17, 2026, Milan, Italy}
\acmYear{2026}
\copyrightyear{2026}
\acmDOI{}
\acmISBN{}

\begin{document}

\title{Neural Realized Measures: Forecast-Supervised Intraday Aggregation for Volatility Forecasting}

\author{Brian Sing Fan Chan}
\affiliation{%
  \department{School of Mathematics and Statistics}
  \institution{The University of Sydney}
  \city{Sydney}
  \state{NSW}
  \country{Australia}
}
\email{brian.chan1@sydney.edu.au}

\author{Jennifer So Kuen Chan}
\affiliation{%
  \department{School of Mathematics and Statistics}
  \institution{The University of Sydney}
  \city{Sydney}
  \state{NSW}
  \country{Australia}
}
\email{jennifer.chan@sydney.edu.au}

\begin{abstract}
High-frequency volatility forecasting typically separates measurement from prediction: intraday observations are first compressed into a daily realized measure using a fixed aggregation rule, and the resulting statistic is then passed to a forecasting model. This conventional pipeline therefore treats measure construction as forecast-agnostic preprocessing. We ask whether the aggregation stage itself can instead be learned from the downstream forecasting objective while preserving the statistical structure of established realized measures. We propose Neural Realized Measures (NRM), a primitive-preserving, forecast-supervised aggregation framework that retains the original measure-specific intraday contributions but replaces uniform aggregation with positive, mean-normalized context-dependent weights. Uniform weights exactly recover the corresponding raw realized measure (RRM), so the classical construction is nested as a special case. Because NRM remains an explicit weighted sum of established primitive contributions, it also retains direct bar-level attribution.

Across SF, HAR, and MIDAS forecasting specifications, NRM is evaluated against corresponding RRM specifications under controlled comparisons. In Heston simulations, NRM reduces RMSE for both contemporaneous and next-day latent integrated variance across all evaluated primitive--regime combinations. On six cryptocurrencies using five-minute data, NRM improves QLIKE in all 54 paired out-of-sample comparisons, with mean gains of 7.23\%, 8.15\%, and 9.22\% under SF, HAR, and MIDAS, respectively. Validation-selected NRM specifications also achieve the lowest mean QLIKE across the six assets among the evaluated approaches. Ablation and SHAP analyses reveal systematic time-of-day and nonlinear local-context effects in the learned allocation.
\end{abstract}


\maketitle

\section{Introduction}

Volatility forecasting commonly begins by compressing high-frequency
observations into a daily realized measure. For measure family $m$ (Eq.~\eqref{eq:m1}-\eqref{eq:m3}), a
classical construction is
\begin{equation}
    R_t^m=\sum_{k=1}^{K}x^m_{t,k},
    \label{eq:classical_measure}
\end{equation}
where $x^m_{t,k}$ denotes the measure-specific bar-level contribution, referred to below as the primitive. Equivalently, the classical construction can be viewed as a weighted aggregation with $w^m_{t,k}=1$ for every intraday contribution. For example, realized variance uses
$x^{\mathrm{RV}}_{t,k}=r_{t,k}^{2}$, whereas bipower variation uses
$x^{\mathrm{BV}}_{t,k}=\frac{\pi}{2}|r_{t,k}||r_{t,k-1}|$. These primitive constructions are not arbitrary summaries of intraday price variation. Realized and range-based measures have been developed through a substantial statistical literature, with foundations in stochastic-process and estimator theory and a strong empirical track record in volatility measurement and forecasting \cite{andersen2003modeling,barndorff2004bipower,garman1980,martens2007range}. Their statistical structure also remains useful in modern neural forecasting: HARNet explicitly bridges the HAR model and a convolutional neural architecture \cite{reisenhofer2022harnet}, while Zhang et al.~\cite{zhang2024intraday} use past intraday realized volatilities as predictors in neural forecasting models. This evidence motivates building neural learning on established realized-measure structure rather than making end-to-end representation learning a prerequisite.

The conventional pipeline nevertheless contains a hidden restriction: the measurement stage is fixed before the forecasting objective is specified. The primitive determines what each bar contributes, but the aggregation rule determines how those contributions are compressed into the daily statistic, and this rule remains forecast-agnostic. In the classical construction, every primitive contribution receives unit weight regardless of candlestick geometry, intraday position, or magnitude relative to nearby bars. We refer to this fixed-aggregation construction as the raw realized measure (RRM). This motivates the central question of the paper: can the aggregation rule be made task-adaptive without discarding the statistical primitive that defines the realized measure?

We address this question with Neural Realized Measures (NRM), a primitive-preserving, forecast-supervised aggregation framework. For measure family $m$,
\begin{equation}
    X_t^m=\sum_{k=1}^{K}w^{m}_{t,k}x^m_{t,k},
    \qquad
    \frac{1}{K}\sum_{k=1}^{K}w^{m}_{t,k}=1.
    \label{eq:intro_nrm}
\end{equation}
The primitive $x^m_{t,k}$ remains fixed; learning operates only through the aggregation weights $w^m_{t,k}$. These weights are positive and estimated from the downstream forecasting objective using the intraday context of each primitive contribution. Setting $w^m_{t,k}=1$ for all $k$ exactly recovers the corresponding RRM. NRM therefore changes where learning enters the forecasting pipeline. Rather than replacing the realized measure with an unrestricted latent representation, it retains the statistically defined primitive and makes only its aggregation task-adaptive. In this sense, NRM occupies a structured middle ground between fixed econometric measurement and end-to-end neural representation learning.

The constraint also gives the learned measure a direct interpretation. Because $X_t^m$ remains an explicit weighted sum of the original primitive contributions, $w^m_{t,k}x^m_{t,k}$ identifies the contribution assigned to bar $k$, while $w^m_{t,k}=1$ provides a natural classical benchmark. The fixed eight-variable STLG representation in Eq.~\eqref{eq:stlg_input} instantiates the context used by the weighting function through range-normalized candlestick geometry, cyclical clock time, a local log contrast, and completed-day summaries. Global variables use only day-$t$ information to forecast day $t+1$. Each NRM specification is compared with its RRM counterpart under the same forecasting setup, isolating the effect of learned aggregation. HARNet, DeepVol, and XGBoost provide flexible direct benchmarks.

The paper makes four main contributions:
\begin{itemize}[leftmargin=*,nosep]
    \item \textbf{Learned aggregation.}
    Conventional realized measures use fixed within-day aggregation rules. We instead treat the aggregation step as part of the forecasting problem, allowing the importance of intraday observations to be learned directly from the forecasting objective.

    \item \textbf{Primitive preservation.}
    We introduce NRM as a general aggregation framework for realized measures with primitive-level additive representations. NRM reweights the underlying primitive contributions using positive, mean-normalized, and context-dependent weights. Uniform weighting recovers the corresponding conventional realized measure exactly, so the original measure is nested as a special case of the proposed framework.

    \item \textbf{Controlled evaluation.}
    We compare each NRM directly with its RRM counterpart while keeping the underlying primitive and forecasting model unchanged. Across several primitive families and SF, HAR, and MIDAS specifications, simulations and cryptocurrency data show that learned aggregation generally improves both latent-volatility estimation and out-of-sample volatility forecasts.

    \item \textbf{Interpretable allocation.}
    Because NRM remains an explicit weighted sum of the original primitive contributions, the learned weights can be inspected directly. Ablation and SHAP analyses show that the model systematically reallocates volatility information according to time of day and nonlinear local market conditions.
\end{itemize}

\section{Related Literature and Positioning}
\label{sec:literature}

NRM builds on realized-measure research that aggregates intraday returns, ranges, and OHLC information into daily variation \cite{andersen2003modeling,garman1980,rogers1991,martens2007range,barndorff2004bipower,barndorff2008kernels}. Recent work also develops alternative high-frequency volatility
primitives and measurement rules that exploit persistent return
distortions \cite{andersen2023pockets}, candlestick geometry
\cite{li2025wicks}, or, more recently, the occupation structure of
the intraday price path \cite{chan2026occupation}, introducing a
distinct path-based characterization of intraday variation. These approaches expand the information encoded in the primitive itself. NRM studies a complementary design dimension: it keeps a given primitive fixed and learns how its intraday contributions are aggregated for forecasting.

HAR and MIDAS capture persistent volatility through heterogeneous or parameterized historical lags \cite{corsi2009har,ghysels2006midas}. In this paper they serve as downstream forecasting heads: MIDAS aggregates information across days, whereas NRM reallocates primitive contributions within a day. Paired RRM--NRM comparisons therefore isolate the effect of learned aggregation while holding the downstream forecasting specification fixed.

Neural volatility forecasting spans both structured extensions of realized-volatility models and more direct mappings from high-frequency inputs to forecasts. HARNet bridges the HAR structure and dilated convolutional networks \cite{reisenhofer2022harnet}; Zhang et al.~\cite{zhang2024intraday} use past intraday realized volatilities as neural predictors; and DeepVol maps high-frequency information through a dilated causal-convolution architecture \cite{morenopino2024deepvol}. NRM takes a complementary route by placing neural learning at the realized-measure construction stage. It assigns context-dependent positive weights to predefined primitive contributions while preserving an explicit daily realized measure.

NRM is also conceptually related to supervised dimension reduction and learned aggregation. Sufficient forecasting uses information from the prediction target to construct low-dimensional predictive indices \cite{fan2017sufficient}, while attention-based multiple-instance learning learns task-dependent weights for aggregating collections of instances \cite{ilse2018attention}. NRM shares the principle that aggregation can be informed by the downstream task, but imposes a stronger structural restriction: rather than pooling unrestricted learned embeddings, it aggregates predefined, statistically interpretable realized-measure primitives. Learning changes their relative allocation but not the primitive being aggregated, and uniform weights exactly recover the corresponding classical realized measure. NRM is therefore best viewed as forecast-supervised statistical aggregation rather than a generic attention architecture.

Recent time-series research emphasizes task-specific inductive bias over architectural scale alone \cite{kim2024cats,tan2024language,wu2025revisiting,eldele2024tslanet}. NRM follows this principle by placing model flexibility at a narrowly defined stage of the pipeline: the aggregation of statistically grounded intraday contributions. The objective is not to replace domain structure with a larger architecture, but to identify which part of that structure benefits from task-adaptive learning.

\section{Neural Realized Measures}
\label{sec:method}

NRM preserves the measure-specific primitive and learns its
forecast-supervised within-day allocation through positive,
mean-normalized weights.

\subsection{Classical intraday contributions}

For day $t$ and intraday bar $k$, let $O_{t,k}$, $H_{t,k}$, $L_{t,k}$, and $C_{t,k}$ denote log open, high, low, and close prices. Define
\begin{equation*}
    r_{t,k}=C_{t,k}-O_{t,k},
    \qquad
    \rho_{t,k}=H_{t,k}-L_{t,k}.
\end{equation*}
We deliberately use within-bar open-to-close returns so all primitives align with the same OHLC interval and no return crosses the UTC-day boundary. The measure family $m \in \{ {\rm RV,BV,GK}\}$ refers to the following 3 measures.
The return-based realized-variance contribution is
\begin{equation}
    x^{\RV}_{t,k}=r_{t,k}^{2}. \label{eq:m1}
\end{equation}
For $k\geq 2$, the bipower-variation contribution is
\begin{equation}
    x^{\BV}_{t,k}=\frac{\pi}{2}|r_{t,k}||r_{t,k-1}|, \label{eq:m2}
\end{equation}
We set $x^{\BV}_{t,1}=0$, so the adjacent-return product does not cross the UTC-day boundary. The Garman--Klass contribution is
\begin{equation}
    x^{\GK}_{t,k}
    =\frac{1}{2}\rho_{t,k}^{2}-(2\log 2-1)r_{t,k}^{2}. \label{eq:m3}
\end{equation}
RRM and NRM are both constructed from these original measure-specific contributions, with
$R_t^m=\sum_k x^m_{t,k}$. Both the controlled simulation and the empirical comparison use RV, BV, and GK.

\subsection{Shape--Time--Local--Global state representation}
\label{subsec:stlg}

The Shape component is motivated by recent candlestick-based volatility research, which shows that the decomposition of an OHLC bar into its open--close body and wick components contains information about the nature of within-bar price variation \cite{li2025wicks}. Strongly directional movements tend to produce relatively long bodies and short wicks, whereas intrabar reversals produce larger wick components. NRM retains the original RV, BV, or GK primitive and uses candlestick geometry as conditioning information for its aggregation weights.

Define the body, upper wick, and lower wick by
\begin{equation*}
    B_{t,k}=|C_{t,k}-O_{t,k}|, U_{t,k}=H_{t,k}-\max\{O_{t,k},C_{t,k}\}, D_{t,k}=\min\{O_{t,k},C_{t,k}\}-L_{t,k}.
\end{equation*}
To keep the range ratios and logarithmic features below well defined, we fix
$\epsilon=10^{-12}$ and use it only as an additive numerical stabilizer in
the STLG feature construction. Their range-normalized {\em Shape} variables are
\begin{equation*}
    BR_{t,k}=\frac{B_{t,k}}{\rho_{t,k}+\epsilon},\quad
    UR_{t,k}=\frac{U_{t,k}}{\rho_{t,k}+\epsilon},\quad
    LR_{t,k}=\frac{D_{t,k}}{\rho_{t,k}+\epsilon}.
\end{equation*}
Normalizing by the high--low range removes the absolute price scale and yields an interpretable decomposition of bar geometry. The body ratio captures directional open--close movement, while the upper- and lower-wick ratios distinguish upward and downward intrabar excursions. These variables are conditioning features rather than separate volatility measures.

Intraday {\em Time} position is represented by
\begin{equation*}
    T_{t,k}=\left[\sin(2\pi k/K),\cos(2\pi k/K)\right].
\end{equation*}
For a centered same-day neighborhood $\mathcal{N}_{k,q}$ excluding bar $k$, define
\begin{equation*}
    \bar{x}^{m,(-k,q)}_{t,k}
    =\frac{1}{|\mathcal{N}_{k,q}|}
      \sum_{\ell\in\mathcal{N}_{k,q}}x^m_{t,\ell},
\end{equation*}
and the {\em Local} volatility contrast
\begin{equation}
    d^m_{t,k}
    =\log\!\left(x^m_{t,k}+\epsilon\right)
    -\log\!\left(\bar{x}^{m,(-k,q)}_{t,k}+\epsilon\right).
    \label{eq:local_contrast}
\end{equation}
Here, $q$ denotes the nominal total width of a centered rolling window before the current bar is excluded. Thus, an interior window contains up to $q-1$ neighboring contributions, split as evenly as possible before and after bar $k$. Near the UTC-day boundaries, the window is truncated and uses only the available same-day neighbors; it never wraps across days. The validation grid uses $q\in\{12,48,72,144\}$ five-minute bars, corresponding to nominal spans of 1, 4, 6, and 12 hours. The value of $q$ is selected on the validation sample as part of the NRM configuration. The {\em Global} component summarizes the completed same-day distribution of primitive contributions. Let
\begin{align*}
    g^{m,\mathrm{med}}_t
    &=\log\!\left(\operatorname{median}x^m_{t,1:K}+\epsilon\right),\\
    g^{m,\mathrm{sd}}_t
    &=\log\!\left(
      \sqrt{\frac{1}{K}\sum_{j=1}^{K}\left(x^m_{t,j}-\bar{x}^m_t\right)^2}
      +\epsilon\right),
\end{align*}
where $\bar{x}^m_t=K^{-1}\sum_{j=1}^{K}x^m_{t,j}$. The fixed aggregation input is
{\small \begin{equation}
    u^{m,STLG}_{t,k}
    =\left[BR_{t,k},UR_{t,k},LR_{t,k},
    \sin\Big(\frac{2\pi k}{K}\Big),\cos\Big(\frac{2\pi k}{K}\Big),d^m_{t,k},
    g^{m,\mathrm{med}}_t,g^{m,\mathrm{sd}}_t\right].
    \label{eq:stlg_input}
\end{equation} }
The primitive $x^m_{t,k}$ determines what is aggregated, while $u^{m,STLG}_{t,k}$ determines its relative weight. The Local component distinguishes an isolated contribution from a generally volatile neighborhood, and the Global variables condition the shared scorer on the day's primitive level and dispersion. Although volume and cross-market turnover may contain additional volatility information \cite{chan2018stockconnect}, STLG deliberately uses only OHLC- and primitive-derived inputs so that RRM--NRM gains can be attributed to learned aggregation rather than an expanded information set.

The Global summaries use all intraday contributions from day $t$, as does the classical daily realized measure itself. They therefore introduce no look-ahead information for forecasting day $t+1$: the complete day-$t$ record is available when the daily measure is constructed, and the summaries are repeated across bars solely to condition the shared score function. In the contemporaneous simulation exercise, the complete day-$t$ record is likewise available when the constructed measure is compared with oracle $IV_t$.

\subsection{Normalized neural aggregation}

\begin{figure*}[t]
    \centering
    \includegraphics[width=0.7\textwidth]{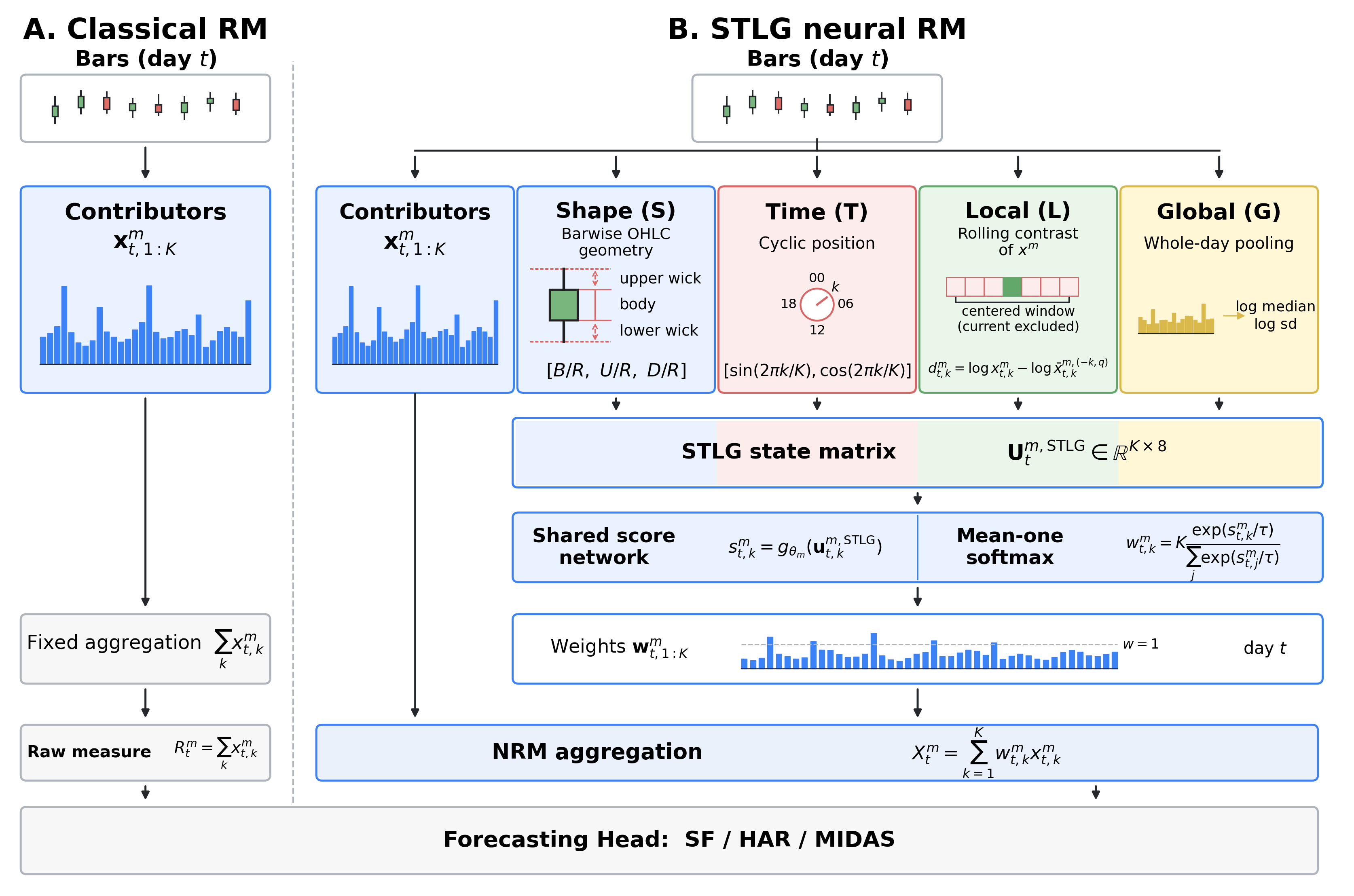}
    \Description{A comparison of the classical fixed aggregation pipeline and the proposed neural realized-measure pipeline with fixed Shape--Time--Local--Global inputs and a same-day local-context benchmark.}
    \caption{Classical realized-measure and NRM pipelines. The classical construction sums fixed primitive contributions. NRM retains the same primitive sequence but forms normalized aggregation weights from the fixed STLG representation and a same-day local benchmark. Its output remains a daily realized measure before entering a downstream forecast model.}
    \label{fig:nrm_pipeline}
\end{figure*}

Figure~\ref{fig:nrm_pipeline} contrasts the classical and NRM pipelines. NRM specifically modifies the within-day aggregation stage; both pipelines subsequently feed a daily measure into the same class of forecast models.

The STLG-conditioned score network produces
\begin{equation}
    s^{m}_{t,k}=g_{\theta_m}(u^{m,STLG}_{t,k}). \label{eq:score}
\end{equation}
The scores are converted into mean-one relative aggregation weights,
\begin{equation}
    w^{m}_{t,k}
    =K\frac{\exp(s^{m}_{t,k}/\tau)}
    {\sum_{j=1}^{K}\exp(s^{m}_{t,j}/\tau)},
    \qquad
    \frac{1}{K}\sum_{k=1}^{K}w^{m}_{t,k}=1,
    \label{eq:nrm_weights}
\end{equation}
where $\tau>0$ controls weight concentration. We fix $\tau=1$ so that concentration is governed by the learned score scale, while validation-selected KL-to-uniform regularization penalizes deviations from classical uniform aggregation. The neural realized measure is
\begin{equation}
X_t^m=\sum_{k=1}^{K}w^{m}_{t,k}x^m_{t,k}.
    \label{eq:nrm_measure}
\end{equation}
Thus $w^{m}_{t,k}=1$ is the classical benchmark and values above (below) one up(down)-weight a bar contribution. A complementary local characterization follows from the softmax parameterization:
\begin{equation}
\frac{\partial X_t^m}{\partial s^m_{t,j}}
=\frac{w^m_{t,j}}{\tau}\left(x^m_{t,j}-\frac{X_t^m}{K}\right).
\label{eq:score_sensitivity}
\end{equation}
Hence increasing a bar's score raises the daily NRM exactly when its primitive contribution exceeds the current NRM-weighted average $X_t^m/K$. The normalization therefore reallocates a fixed total weight mass rather than introducing an unrestricted scale multiplier. Because the primitive remains explicit, the contribution of bar $k$ remains $w^m_{t,k}x^m_{t,k}$, preserving primitive-level attribution.

\subsection{Estimation and paired-comparison protocol}
\label{subsec:paired_protocol}

\paragraph{Network structures.}
Neural networks estimate both
the aggregation weight $w^{m}_{t,k}$ in Eq.~\eqref{eq:nrm_weights} for NRM in Eq.~\eqref{eq:nrm_measure} and the forecast model parameters. To quantify departure from equal-weight aggregation, define the implied simplex weights $\alpha^m_{t,k}=w^m_{t,k}/K$, the uniform allocation $u_k=1/K$, and the daily divergence
\begin{equation}
D_t^m:=D_{\mathrm{KL}}(\alpha_t^m\Vert u)
=\frac{1}{K}\sum_{k=1}^K w^m_{t,k}\log w^m_{t,k}.
\label{eq:daily_kl}
\end{equation}
For a mini-batch $\mathcal B$ of daily observations, the network loss penalizes the average daily divergence:
\begin{equation}
  \mathrm{Loss} = {\rm QLIKE} + \lambda\mathcal{L}_{\mathrm{KL}}^{m},
  \qquad
  \mathcal{L}_{\mathrm{KL}}^{m}
  =\frac{1}{|\mathcal B|}\sum_{t\in\mathcal B} D_t^m.
  \label{eq:loss}
\end{equation}

\begin{proposition}[KL-controlled departure from classical aggregation]
\label{prop:kl_departure}
For a fixed day $t$ and measure family $m$, let
$\Delta_t^m=\max_k x^m_{t,k}-\min_k x^m_{t,k}$ and let $D_t^m$ be defined by Eq.~\eqref{eq:daily_kl}. Then
\begin{equation}
\frac{\left|X_t^m-R_t^m\right|}{K}
\le \Delta_t^m\sqrt{\frac{1}{2}D_t^m}.
\label{eq:kl_departure_bound}
\end{equation}
Equivalently,
$\left|X_t^m-R_t^m\right|
\le K\Delta_t^m\sqrt{D_t^m/2}$.
In particular, $D_t^m=0$ if and only if $w^m_{t,k}=1$ for all $k$, in which case $X_t^m=R_t^m$. If $\Delta_t^m=0$, then $X_t^m=R_t^m$ for any admissible allocation.
\end{proposition}

\begin{proof}
By Eq.~\eqref{eq:nrm_weights}, $w^m_{t,k}>0$ and
$\sum_{k=1}^K w^m_{t,k}=K$. Hence
$\alpha^m_{t,k}=w^m_{t,k}/K$ is a probability vector, as is
$u_k=1/K$. Moreover,
\[
\frac{X_t^m}{K}
=\sum_{k=1}^K\alpha^m_{t,k}x^m_{t,k},
\qquad
\frac{R_t^m}{K}
=\sum_{k=1}^K u_kx^m_{t,k}.
\]
Let
$c_t^m=(\max_k x^m_{t,k}+\min_k x^m_{t,k})/2$.
Since $\sum_k(\alpha^m_{t,k}-u_k)=0$, subtracting the common
center $c_t^m$ does not change the difference:
\begin{align*}
\frac{|X_t^m-R_t^m|}{K}
&=\left|
\sum_{k=1}^K(\alpha^m_{t,k}-u_k)(x^m_{t,k}-c_t^m)
\right| \\
&\le
\frac{\Delta_t^m}{2}
\sum_{k=1}^K|\alpha^m_{t,k}-u_k| \\
&=
\Delta_t^m\,\mathrm{TV}(\alpha_t^m,u),
\end{align*}
where
$\mathrm{TV}(\alpha_t^m,u)
=\frac12\|\alpha_t^m-u\|_1$.
Pinsker's inequality gives
\[
\mathrm{TV}(\alpha_t^m,u)
\le
\sqrt{\frac12D_{\mathrm{KL}}(\alpha_t^m\Vert u)}
=
\sqrt{\frac12D_t^m},
\]
which proves Eq.~\eqref{eq:kl_departure_bound}.

For completeness, the identity in Eq.~\eqref{eq:daily_kl} follows directly from
$\alpha^m_{t,k}=w^m_{t,k}/K$:
\[
D_{\mathrm{KL}}(\alpha_t^m\Vert u)
=
\sum_{k=1}^K\frac{w^m_{t,k}}{K}
\log\!\left(\frac{w^m_{t,k}/K}{1/K}\right)
=
\frac1K\sum_{k=1}^K w^m_{t,k}\log w^m_{t,k}.
\]
KL divergence is zero if and only if its two probability vectors coincide,
so $D_t^m=0$ exactly when $\alpha_t^m=u$, equivalently
$w^m_{t,k}=1$ for every $k$; Eq.~\eqref{eq:nrm_measure} then gives
$X_t^m=R_t^m$. Finally, if $\Delta_t^m=0$, all primitive contributions
are equal, and the mean-one constraint
$\sum_k w^m_{t,k}=K$ gives $X_t^m=R_t^m$ for any admissible allocation.
\end{proof}

The bound shows that departure from the corresponding RRM is jointly governed by within-day primitive dispersion and divergence from uniform allocation. The KL penalty therefore regularizes the same allocation divergence that appears in Proposition~\ref{prop:kl_departure}. The tuning parameter $\lambda$ is selected together with the other network hyperparameters. QLIKE is given by
\begin{equation}
    \mathrm{QLIKE}(V_t,\widehat V_t)
    =\frac{V_t}{\widehat V_t}-\log\!\left(\frac{V_t}{\widehat V_t}\right)-1, \label{eq:QLIKE}
\end{equation}
The network outputs the log-variance forecast $\widehat{y}_t=\log \widehat{V}_t$, while $y_t=\log V_t$ denotes the corresponding target.

The score network in Eq.~\eqref{eq:score} generates the aggregation weights in Eq.~\eqref{eq:nrm_weights} and is estimated separately for each asset, measure family $m$, and downstream forecasting specification in $\{\mathrm{SF},\mathrm{HAR},\mathrm{MIDAS}\}$ (Section~\ref{sec:forecast_specs}). The forecasting-specification index is suppressed in $X_t^m$ for notational simplicity, even though the NRM aggregation weights and downstream forecasting parameters are jointly estimated under Eq.~\eqref{eq:loss}. Accordingly, $X_t^m$ is forecasting-specification-specific: its learned allocation adapts directly to the corresponding forecasting objective rather than requiring one fixed NRM series to serve all downstream specifications. When the equal-weight RRM $R_t^m$ in Eq.~\eqref{eq:classical_measure} is used, the NRM aggregation layer is omitted.
\paragraph{The protocol.}
\begin{enumerate}
    \item fit each neural candidate on the training sample using QLIKE and the KL-to-uniform penalty;
    \item early-stop each candidate using validation log-MSE; for each candidate NRM representation, the corresponding
    forecasting-head coefficients are re-estimated using training observations only;
    \item select one common specification using mean validation MSE on the variance scale across seeds 11, 22, and 33;
    \item retain the three independently trained fits of that selected specification, evaluate each once on the held-out test period, and report test losses averaged across the seeds. No test observation is used for fitting, specification selection, or seed selection.
\end{enumerate}
The corresponding RRM specification uses the same forecasting and
evaluation setup but retains uniform intraday weights. This paired design evaluates the incremental value of learned aggregation while holding the downstream forecasting specification fixed; a separate NRM is estimated for each forecasting specification.

\par\noindent\textit{Implementation and computational details.}\enspace
The aggregation weight layer for NRM score in Eq~\eqref{eq:score} uses GELU activation, dropout 0.1, and hidden dimensions selected from $[8,4]$ and $[16,8]$. The validation grid considers 1-, 4-, 6-, and 12-hour centered same-day local windows, corresponding to 12, 48, 72, and 144 five-minute bars. We fix softmax temperature at $\tau=1$ and select $\lambda$ for the KL-to-uniform strength from $\{3\!\times\!10^{-4},10^{-3},3\!\times\!10^{-3},10^{-2},3\!\times\!10^{-2}\}$; this regularizer directly governs how strongly the fitted weights may depart from the classical uniform allocation. Neural candidates are trained with AdamW, batch size 16, a maximum of 200 epochs, and validation-based early stopping with patience 20; and the
learning rate $3\times10^{-3}$. Experiments were executed on an Ubuntu 22.04 CPU server with Intel Xeon Platinum 8480C processors and approximately 2 TiB of system memory.

\section{Study Design}
\label{sec:design}

\subsection{Controlled simulation}
\label{subsec:simulation_design}

The simulation follows a Heston stochastic-volatility design~\cite{heston1993closed} with the event stresses of Li et al.~\cite{li2025wicks}, but draws event locations uniformly to prevent the time-aware NRM from learning deterministic timing. Continuous diffusion provides a baseline, an isolated jump tests sparse contamination, and a gradual jump tests a multi-bar directional disturbance. No separate conventional microstructure-noise process is added. The efficient log price satisfies
\begin{equation*}
    dX_t=\mu\,dt+\sigma_t\,dW_{1,t}+dJ_t,
\end{equation*}
\begin{equation*}
    d\sigma_t^2=\kappa(\theta-\sigma_t^2)dt+\eta\sigma_t\,dW_{2,t},
    \qquad \mathrm{Corr}(dW_{1,t},dW_{2,t})=\rho,
\end{equation*}
where $J_t$ is a compound-Poisson price-jump process. Observed prices additionally include a short-lived persistent component,
\begin{equation*}
    X_t^{\mathrm{obs}}=X_t+H_t,
\end{equation*}
which generates the persistent displacement used in the gradual-jump stress regime. The reported regimes are:
\begin{enumerate}
    \item \textbf{Continuous}: stochastic volatility without price jumps or persistent noise;
    \item \textbf{Jump}: stochastic volatility with compound-Poisson price jumps;
    \item \textbf{Gradual jump}: a temporary persistent displacement following a fundamental-price shift;
\end{enumerate}
The annualized parameterization is $X_0=\log(1200)$, $\mu=0.05$,
$\kappa=5$, $\theta=0.0225$, $\eta=0.4$, $\rho=-1/\sqrt{2}$, jump
intensity $\lambda=1/5$, and jump-size standard deviation $0.9\%$.
The Euler step is $\Delta t=5/(252\times24\times60)$, yielding 288
five-minute bars per simulated day. Each regime contains 2,373 simulated
days and is split chronologically into 60\% training, 20\% validation,
and 20\% test samples. The gradual-jump regime imposes one event
per simulated day as a conditional stress. Both exercises use the same SF specification and validation
criterion; NRM is selected from the aggregation grid, whereas RRM uses
uniform intraday weights. Results are aggregated across ten Monte Carlo
seeds, varying both the simulated path and neural initialization.

Both simulation targets measure continuous integrated variance; jumps and temporary displacements enter observed OHLC bars but not the oracle. The contemporaneous target is
\begin{equation}
    \IV_{d}^{\mathrm{oracle}}
    =\int_{d}^{d+1}\sigma_s^2\,ds, \label{eq:IV}
\end{equation}
which evaluates how closely the constructed daily measure aligns with same-day latent integrated variance. The next-day forecasting target is $\IV_{d+1}^{\mathrm{oracle}}$ which evaluates next-day predictive information. For each target, the same output specification uses either the RRM measure $R_d^m$ or its corresponding NRM measure $X_d^m$, constructed from the fixed STLG specification; the two exercises assess contemporaneous measurement and forecasting content, respectively.

\subsection{Cryptocurrency data and sample splits}

The empirical analysis uses five-minute OHLC bars from Binance spot price indices for BNB, BTC, DOGE, ETH, SOL, and XRP. These indices aggregate prices from major centralized exchanges, providing a cross-venue reference price that reduces sensitivity to individual-venue liquidity, dislocations, and bid--ask spreads, although it does not eliminate all microstructure noise. Continuous trading permits a common UTC-day definition without overnight or session conventions, but limits external validity for traditional markets. A complete day contains $K=288$ bars; incomplete days are excluded. Samples begin from each asset's first complete day after January 1, 2020 and end on May 31, 2026. Validation covers June 2023--May 2024 and testing June 2024--May 2026, with splits defined by target date.

Latent integrated variance is not observed in market data. The empirical target {\em for all models} is the next-day log realized variance,
    $y_{t+1}=\log R^{\rm RV}_{t+1}$,
where $V_t=
R^{\rm RV}_{t+1}$ in Eq.~\eqref{eq:classical_measure} uses the next day's five-minute returns. Models forecast log realized variance, and variance predictions are $\widehat{V}_{t+1}=\exp(\widehat y_{t+1})$ without smearing or clipping. RMSE, MAE, and QLIKE are evaluated on the original scale without log transformation.

\subsection{Forecasting specifications and flexible benchmarks} \label{sec:forecast_specs}

We consider three downstream forecasting specifications SF, HAR and MIDAS. In the joint architecture, each specification serves as the forecasting head applied after construction of the daily measure. For each specification, the daily variance input is $Q_t=R_t^m$ for RRM and $Q_t=X_t^m$ for NRM, $m \in \{{\rm RV,BV,GK}\}$. The three specifications are
\begin{eqnarray*}
 \mbox{SF:} \   y_{t+1} & = & \beta_0+\beta_1\log Q_t+\varepsilon_{t+1}. \\
 \mbox{HAR:} \ y_{t+1} & = & \beta_0+\beta_dD_t(Q)+\beta_wW_t(Q)+\beta_mM_t(Q)+\varepsilon_{t+1}, \\
 \mbox{MIDAS:} \  y_{t+1} & = & \beta_0+\beta_1\sum_{j=0}^{J-1}b_j(\vartheta)\log Q_{t-j}+\varepsilon_{t+1}, 
\end{eqnarray*}
where
\begin{equation*}
    D_t(Q)=\log Q_t,\hspace{2mm}
    W_t(Q)=\frac{1}{5}\sum_{j=0}^{4}\log Q_{t-j}, \hspace{2mm} M_t(Q)=\frac{1}{22}\sum_{j=0}^{21}\log Q_{t-j}.
\end{equation*}
and the MIDAS specification \citep{ghysels2006midas} replaces the fixed weekly and monthly averages with a normalized distributed-lag function, where $J$ is the lag length and the weights satisfy $\sum_{j=0}^{J-1}b_j(\vartheta)=1$.

{\em Three flexible benchmarks} are added to the 18 specifications: 3 forecasting specifications $\times$ RRM/NRM $\times$ 3 measures (Table~\ref{tab:model_spec_performance}). HARNet learns nonlinear hierarchical transformations of realized-volatility histories. The DeepVol benchmark follows the dilated causal-convolution design of \cite{morenopino2024deepvol} with a capacity-controlled specification for the available daily forecasting sample. XGBoost \citep{chen2016xgboost} uses daily, weekly, and monthly HAR volatility factors together with lagged signed and absolute daily returns as inputs, providing a non-neural nonlinear benchmark. Forecasts are daily, and each daily realized measure is constructed from 288 five-minute bars. These benchmarks provide flexible direct-forecasting references outside the SF, HAR, and MIDAS specifications.

\subsection{Model selection and evaluation}

For every reported specification, the hyperparameter configuration is selected by mean validation MSE averaged across repeats using seeds 11, 22, and 33. This common criterion provides a consistent selection rule and avoids using QLIKE for both parameter optimization and configuration selection. For each selected configuration, three independently trained fits are obtained using the same three seeds. Each fitted model forecasts the full test sample, and test RMSE, MAE, and QLIKE are averaged across the three repeats and reported in Table~\ref{tab:model_spec_performance}, with performance summaries in Tables~\ref{tab:paired_raw_nrm} and \ref{tab:selected_asset_results}.

The primary comparison pairs each NRM specification with its corresponding RRM specification while holding the downstream forecasting specification fixed. This design evaluates the incremental value of learned aggregation; parameter counts need not be equal because the NRM aggregation layer is omitted for RRM.
QLIKE is included in the comparison because it supports robust volatility-forecast comparison when latent variance is evaluated using a noisy proxy \citep{patton2011volatility}.
RMSE and MAE are also reported on the variance scale.
Simulation tables report relative RMSE and MAE,
\begin{equation}
    rRMSE=\frac{RMSE(\mathrm{NRM}\mbox{-}m)}{RMSE(\mathrm{RRM}\mbox{-}m)},
    \qquad
    rMAE=\frac{MAE(\mathrm{NRM}\mbox{-}m)}{MAE(\mathrm{RRM}\mbox{-}m)}. \label{eq:rRMSE}
\end{equation}
Values below one indicate improvement from neural aggregation. Forecast-comparison inference is conducted only after validation selection. For each asset--primitive--forecasting-specification comparison, the selected RRM and NRM specifications generate date-aligned seed-level forecasts, and losses are first averaged across seeds by date. Following the pooled forecast-comparison construction in \cite{cotturo2026multifactor}, we then average the RRM-minus-NRM loss differentials across the available asset--primitive pairs within each forecasting-specification--metric family on every test date. The pooled Diebold--Mariano test \citep{diebold1995comparing} is applied to the resulting time series of cross-sectional mean differentials. Long-run variance is estimated using a Bartlett HAC estimator with automatic lag selection, and the Harvey--Leybourne--Newbold finite-sample correction \citep{harvey1997testing} is applied. Positive statistics favor NRM. Table~\ref{tab:paired_raw_nrm} reports the pooled statistics and one-sided unadjusted $p$-values. Statistical testing is conducted only after validation-based specification selection: candidate configurations are not individually tested on the test sample, and no model choice is based on test-set performance.

\section{Results}
\label{sec:results}

\subsection{Controlled simulation evidence}

Table~\ref{tab:simulation_ratio_results} reports absolute and relative RMSE for contemporaneous $\IV_t$ and next-day $\IV_{t+1}$. Jump and gradual-jump movements enter observed OHLC contributions but not the continuous-IV oracle. Reporting absolute errors beside ratios separates lower NRM error from inflation of the RRM denominator.
\begin{table}[t]
\centering
\tiny
\setlength{\tabcolsep}{1.35pt}
\renewcommand{\arraystretch}{0.84}
\caption{Simulation absolute and relative RMSE}
\label{tab:simulation_ratio_results}
\resizebox{\columnwidth}{!}{%
\begin{tabular}{ll|rrr|rrr|rrr}
\toprule
Target & Measure & \multicolumn{3}{c|}{Continuous} & \multicolumn{3}{c|}{Jump} & \multicolumn{3}{c}{Gradual jump} \\
 &  & RRM & NRM & $r\mathrm{RMSE}$ & RRM & NRM & $r\mathrm{RMSE}$ & RRM & NRM & $r\mathrm{RMSE}$ \\
\midrule
\multirow{3}{*}{$\mathrm{IV}_t$} & RV & 0.981 & 0.607 & 0.619 & 4.820 & 0.665 & 0.138 & 7.658 & 1.433 & 0.187 \\
 & BV & 1.118 & 0.857 & 0.767 & 1.171 & 0.841 & 0.718 & 3.800 & 2.601 & 0.685 \\
 & GK & 0.400 & 0.391 & 0.976 & 1.266 & 0.362 & 0.286 & 1.624 & 0.376 & 0.232 \\
\midrule
\multirow{3}{*}{$\mathrm{IV}_{t+1}$} & RV & 1.606 & 1.476 & 0.919 & 4.784 & 1.386 & 0.290 & 7.245 & 1.687 & 0.233 \\
 & BV & 1.675 & 1.644 & 0.981 & 1.664 & 1.504 & 0.904 & 3.741 & 2.476 & 0.662 \\
 & GK & 1.372 & 1.294 & 0.943 & 1.754 & 1.164 & 0.663 & 1.926 & 1.137 & 0.590 \\
\bottomrule
\end{tabular}%
}
{
\raggedright
\scriptsize Notes: RRM and NRM columns report absolute RMSE
in units of $10^{-5}$. $r\mathrm{RMSE}$ is defined in Eq.\eqref{eq:rRMSE}.
Values below one favor NRM. RMSE are averaged across simulations (by seeds) before ratios rRMSE are formed. Reporting RMSEs distinguishes improvement in NRM from inflation of the RRM in the  denominator.
\par
}
\end{table}

NRM lowers RMSE in every primitive--regime combination. Gains are modest under continuous diffusion and larger under event stress, where OHLC contributions contain movements excluded from the oracle. The small RV and GK ratios reflect both lower NRM error and larger RRM error under estimand mismatch, not near-perfect latent-variance recovery. Next-day gains are smaller but broad. The simulation diagnoses finite-sample measurement and forecasting content, not unbiasedness or consistency.

\subsection{Empirical forecasting performance}

\begin{table}[t]
\centering
\scriptsize
\setlength{\tabcolsep}{3.0pt}
\caption{Out-of-Sample Performance of NRM over RRM by Forecasting Specification and Metric across 18 Specifications (6 Assets $\times$ 3 Measure Families)}
\label{tab:paired_raw_nrm}
\resizebox{\columnwidth}{!}{%
\begin{tabular}{llrrrrr}
\toprule
Specification& Metric & Mean gain (\%) & Median gain (\%) & NRM wins & DM & $p$ \\
\midrule
SF & QLIKE & 7.23 & 6.99 & 18/18 & 4.06 & $<0.001$ \\
 & RMSE & 3.35 & 3.41 & 17/18 & 1.81 & 0.035 \\
 & MAE & 5.13 & 4.40 & 18/18 & 2.05 & 0.020 \\
\addlinespace[1.5pt]
HAR & QLIKE & 8.15 & 8.08 & 18/18 & 3.08 & 0.001 \\
 & RMSE & 1.30 & 1.20 & 17/18 & 2.70 & 0.003 \\
 & MAE & 1.39 & 1.33 & 15/18 & 1.69 & 0.046 \\
\addlinespace[1.5pt]
MIDAS & QLIKE & 9.22 & 10.22 & 18/18 & 3.20 & $<0.001$ \\
 & RMSE & 1.54 & 1.60 & 17/18 & 2.83 & 0.002 \\
 & MAE & 1.37 & 1.40 & 16/18 & 1.54 & 0.062 \\
\bottomrule
\end{tabular}%
}
{
\raggedright
\footnotesize Notes: Gains are percentages averaged over 18 specifications.
DM and $p$ are the pooled Diebold--Mariano statistic and one-sided unadjusted $p$-value; positive DM values favor NRM.
\par
}
\end{table}

\begin{table}[t]
\centering
\tiny
\setlength{\tabcolsep}{3.5pt}
\caption{Out-of-Sample QLIKE Performance after Validation-MSE Selection across Assets}
\label{tab:selected_asset_results}
\resizebox{\columnwidth}{!}{%
\begin{tabular}{lrrrrr}
\toprule
Asset & Selected RRM & Selected NRM & HARNet & DeepVol & XGB \\
\midrule
BTC & 0.301 & \textbf{0.278} & 0.326 & 0.335 & \underline{0.294} \\
ETH & 0.272 & \textbf{0.260} & 0.293 & 0.328 & \underline{0.265} \\
SOL & \underline{0.253} & \textbf{0.246} & 0.272 & 0.272 & 0.256 \\
BNB & \underline{0.348} & \textbf{0.342} & 0.360 & 0.381 & 0.375 \\
XRP & \underline{0.560} & \textbf{0.541} & 0.648 & 0.632 & 0.698 \\
DOGE & \underline{0.648} & \textbf{0.579} & 0.751 & 0.689 & 0.666 \\
\midrule
Mean & \underline{0.397} & \textbf{0.374} & 0.442 & 0.439 & 0.426 \\
\bottomrule
\end{tabular}%
}
{
\raggedright
\footnotesize Notes:
Lower values are better; bold indicates the lowest value and underline indicates the second-lowest value in each row.
\par
}
\end{table}

\noindent\textbf{Paired RRM--NRM comparisons.}
Table~\ref{tab:paired_raw_nrm} shows broad improvements across all three forecasting specifications. NRM wins all 54 QLIKE comparisons, with mean gains of 7.23\%, 8.15\%, and 9.22\% under SF, HAR, and MIDAS, respectively, and all pooled QLIKE tests are significant. RMSE improvements occur in 17 of 18 pairs under each forecasting specification. MAE gains are also broad, although the pooled MIDAS test is not significant at 5\% ($p=0.062$). Because each pair holds the downstream forecasting specification and data setup fixed, these results support improved predictive content from learned aggregation rather than gains produced by changing the downstream forecasting structure.

\noindent\textbf{Flexible benchmark comparison.}
Table~\ref{tab:selected_asset_results} compares validation-selected RRM and NRM specifications with HARNet, the DeepVol model, and XGBoost. NRM has the lowest mean QLIKE and improves on the validation-selected RRM specification for all six assets. The magnitude of the gain is smallest for BNB and larger for BTC and DOGE.

Performance remains heterogeneous across assets and metrics, and the strongest competing benchmark differs across markets. Across the six assets, however, the validation-selected NRM specifications attain the lowest mean QLIKE among the compared approaches. The analysis does not identify which market characteristics explain the remaining cross-asset differences.

Table~\ref{tab:model_spec_performance} reports the full asset-by-specification losses underlying the compact comparison. It is retained in the main text to show heterogeneity across assets, forecasting specifications, and realized-measure primitives. The table reports test performance for specifications selected on the validation set, rather than the best test-set performance among candidate specifications.

\begin{table*}[t]
\centering
\tiny
\setlength{\tabcolsep}{1.05pt}
\renewcommand{\arraystretch}{0.86}
\caption{Out-of-Sample Forecasting Performance across Forecasting Specifications and Realized Measures}
\label{tab:model_spec_performance}
\resizebox{\textwidth}{!}{%
\begin{tabular}{ll|ccc|ccc|ccc|ccc|ccc|ccc|c|c|c}
\toprule
\multirow{2}{*}{Ticker}
& \multirow{2}{*}{Metric}
& \multicolumn{3}{c|}{SF}
& \multicolumn{3}{c|}{SF-NRM}
& \multicolumn{3}{c|}{HAR}
& \multicolumn{3}{c|}{HAR-NRM}
& \multicolumn{3}{c|}{MIDAS}
& \multicolumn{3}{c|}{MIDAS-NRM}
& \multirow{2}{*}{HARNet}
& \multirow{2}{*}{DeepVol}
& \multirow{2}{*}{XGB} \\
\cmidrule(lr){3-5}
\cmidrule(lr){6-8}
\cmidrule(lr){9-11}
\cmidrule(lr){12-14}
\cmidrule(lr){15-17}
\cmidrule(lr){18-20}
& & RV & BV & GK & RV & BV & GK & RV & BV & GK
& RV & BV & GK & RV & BV & GK & RV & BV & GK & & & \\
\midrule

\multirow{3}{*}{\textbf{BTC}}
& QLIKE & 0.316 & 0.315 & 0.343 & 0.292 & 0.295 & 0.300 & 0.311 & 0.301 & 0.346 & \underline{0.284} & \textbf{0.278} & 0.291 & 0.323 & 0.312 & 0.337 & 0.286 & 0.291 & 0.293 & 0.326 & 0.335 & 0.294 \\
& RMSE & 0.782 & 0.788 & 0.775 & 0.751 & 0.759 & 0.751 & 0.750 & 0.747 & 0.752 & 0.729 & 0.729 & \underline{0.727} & 0.750 & 0.746 & 0.751 & \textbf{0.726} & 0.728 & 0.732 & 0.781 & 0.763 & 0.745 \\
& MAE & 0.353 & 0.360 & 0.345 & 0.337 & 0.340 & 0.337 & 0.321 & 0.326 & 0.323 & 0.319 & \textbf{0.316} & 0.319 & 0.321 & 0.326 & 0.323 & \underline{0.318} & 0.319 & 0.322 & 0.327 & 0.327 & 0.321 \\
\midrule

\multirow{3}{*}{\textbf{ETH}}
& QLIKE & 0.281 & 0.277 & 0.275 & 0.262 & 0.260 & 0.268 & 0.280 & 0.272 & 0.271 & \underline{0.258} & 0.260 & 0.259 & 0.297 & 0.289 & 0.285 & 0.265 & 0.273 & \textbf{0.250} & 0.293 & 0.328 & 0.265 \\
& RMSE & 1.509 & 1.505 & 1.534 & 1.472 & 1.462 & 1.496 & 1.465 & 1.460 & 1.457 & 1.439 & 1.442 & 1.438 & 1.475 & 1.469 & 1.468 & \underline{1.431} & 1.442 & \textbf{1.415} & 1.524 & 1.480 & 1.448 \\
& MAE & 0.674 & 0.678 & 0.687 & 0.649 & 0.653 & 0.677 & 0.624 & 0.630 & 0.641 & \underline{0.615} & 0.621 & 0.640 & 0.624 & 0.627 & 0.639 & \textbf{0.614} & 0.624 & 0.624 & 0.643 & 0.629 & 0.619 \\
\midrule

\multirow{3}{*}{\textbf{SOL}}
& QLIKE & 0.265 & 0.262 & 0.258 & 0.246 & 0.253 & 0.247 & 0.263 & 0.258 & 0.253 & \textbf{0.242} & 0.249 & 0.243 & 0.263 & 0.258 & 0.256 & 0.245 & 0.253 & \underline{0.242} & 0.272 & 0.272 & 0.256 \\
& RMSE & 3.009 & 3.013 & 2.925 & 2.860 & 3.049 & 2.820 & 2.866 & 2.867 & 2.847 & 2.798 & 2.838 & \underline{2.784} & 2.848 & 2.842 & 2.839 & 2.792 & 2.799 & \textbf{2.777} & 3.053 & 2.897 & 2.918 \\
& MAE & 1.125 & 1.141 & 1.118 & 1.081 & 1.125 & 1.061 & 1.014 & 1.030 & 1.036 & \underline{1.001} & 1.012 & 1.006 & 1.012 & 1.024 & 1.027 & 1.008 & 1.005 & \textbf{0.997} & 1.029 & 1.053 & 1.050 \\
\midrule

\multirow{3}{*}{\textbf{BNB}}
& QLIKE & 0.336 & 0.339 & 0.372 & \textbf{0.318} & \underline{0.326} & 0.355 & 0.350 & 0.348 & 0.393 & 0.334 & 0.342 & 0.361 & 0.357 & 0.354 & 0.397 & 0.326 & 0.343 & 0.358 & 0.360 & 0.381 & 0.375 \\
& RMSE & 5.478 & 5.615 & 5.191 & 5.163 & 5.188 & 5.160 & 5.189 & 5.205 & 5.169 & 5.172 & 5.223 & 5.165 & 5.164 & 5.169 & 5.160 & \textbf{5.149} & 5.169 & \underline{5.156} & 5.764 & 5.163 & 5.166 \\
& MAE & 0.707 & 0.720 & 0.632 & 0.616 & 0.627 & 0.586 & 0.615 & 0.621 & 0.595 & 0.589 & 0.613 & \textbf{0.579} & 0.614 & 0.617 & 0.596 & 0.586 & 0.602 & \underline{0.584} & 0.734 & 0.603 & 0.600 \\
\midrule

\multirow{3}{*}{\textbf{XRP}}
& QLIKE & 0.622 & 0.610 & 0.545 & 0.541 & 0.541 & \textbf{0.497} & 0.651 & 0.628 & 0.560 & 0.549 & 0.552 & \underline{0.502} & 0.650 & 0.629 & 0.579 & 0.536 & 0.560 & 0.513 & 0.648 & 0.632 & 0.698 \\
& RMSE & 5.999 & 5.936 & 6.098 & 5.715 & 5.733 & 5.769 & 5.786 & 5.777 & 5.790 & 5.720 & 5.729 & \textbf{5.703} & 5.803 & 5.796 & 5.796 & \underline{5.705} & 5.730 & 5.713 & 6.173 & 5.787 & 5.784 \\
& MAE & 1.444 & 1.445 & 1.477 & 1.381 & 1.397 & 1.435 & 1.349 & 1.355 & 1.381 & 1.355 & 1.356 & 1.396 & 1.352 & 1.356 & 1.367 & \underline{1.345} & 1.365 & 1.380 & 1.447 & 1.370 & \textbf{1.335} \\
\midrule

\multirow{3}{*}{\textbf{DOGE}}
& QLIKE & 0.634 & 0.631 & 0.675 & \textbf{0.579} & \underline{0.580} & 0.610 & 0.660 & 0.648 & 0.704 & 0.598 & 0.593 & 0.623 & 0.674 & 0.655 & 0.721 & 0.589 & 0.611 & 0.644 & 0.751 & 0.689 & 0.666 \\
& RMSE & 16.710 & 16.478 & 16.204 & 16.140 & 16.065 & \underline{16.018} & 16.147 & 16.124 & 16.093 & 16.075 & 16.048 & 16.019 & 16.085 & 16.072 & 16.068 & 16.033 & 16.041 & \textbf{16.009} & 18.033 & 16.051 & 16.041 \\
& MAE & 1.972 & 1.931 & 1.839 & 1.836 & 1.821 & 1.757 & 1.809 & 1.809 & 1.765 & 1.786 & 1.789 & \underline{1.728} & 1.806 & 1.801 & 1.769 & 1.786 & 1.798 & \textbf{1.727} & 2.099 & 1.788 & 1.747 \\
\bottomrule
\end{tabular}%
}
{
\raggedright
\footnotesize Notes: The table reports ticker-level out-of-sample losses after selecting each model specification by validation MSE. RMSE and MAE entries are reported in units of $10^{-3}$; QLIKE is reported on its original scale. Lower values are better. For SF, HAR, and MIDAS, each subcolumn uses the corresponding daily RRM  family. Their NRM counterparts use the corresponding validation-selected NRM daily series constructed from the same primitive family. HARNet, DeepVol, and XGB are direct benchmark models and therefore do not have realized-measure-family subcolumns. Bold entries indicate the best model within each cryptocurrency-loss row, and underlined entries indicate the second-best model.
\par
}
\end{table*}

\subsection{Feature-group ablation}

Table~\ref{tab:feature_ablation} reports controlled leave-one-group-out
ablations of the HAR--NRM specifications. Removing Time and Local produces
the largest average deterioration, increasing QLIKE by 4.54\% and 4.01\%,
respectively, followed by Global at 2.54\%. Shape has the smallest and most
measure-dependent effect. The main gains therefore arise from intraday
timing and local context, with additional information from the completed-day
distribution.

\begin{table}[t]
\centering
\footnotesize
\setlength{\tabcolsep}{3.5pt}
\caption{Controlled HAR--NRM Leave-One-Feature-Group-Out Ablation}
\label{tab:feature_ablation}
\begin{tabular}{lrrrrrrrr}
\toprule
& \multicolumn{4}{c}{$\Delta$QLIKE} & \multicolumn{4}{c}{$\Delta$RMSE} \\
\cmidrule(lr){2-5}\cmidrule(lr){6-9}
Measure & Drop S & Drop T & Drop L & Drop G & Drop S & Drop T & Drop L & Drop G \\
\midrule
RV & +1.97\% & +3.32\% & +1.59\% & +3.16\% & +0.24\% & +1.04\% & +0.46\% & +0.24\% \\
BV & -0.48\% & +3.76\% & +2.87\% & +0.62\% & -0.24\% & +0.79\% & +0.73\% & -0.23\% \\
GK & +1.84\% & +6.54\% & +7.56\% & +3.84\% & +0.88\% & +1.36\% & +0.49\% & +0.60\% \\
\midrule
Overall & +1.11\% & +4.54\% & +4.01\% & +2.54\% & +0.29\% & +1.06\% & +0.56\% & +0.20\% \\
\bottomrule
\end{tabular}
\begin{minipage}{0.98\columnwidth}
\scriptsize
\textit{Notes:} For each asset--measure pair, the full STLG specification is selected by mean validation MSE across the main-experiment seeds. Each ablated model inherits that complete hyperparameter configuration and is retrained after changing only the included feature groups. Entries are mean seed-paired percentage changes in test loss, first averaged across matched seeds within each asset--measure pair and then across assets. Positive values indicate worse performance after removing a feature group.
\end{minipage}
\end{table}

\section{Mechanism and Interpretation}
\label{sec:mechanism}

Because NRM retains explicit weights on established primitive contributions, its learned allocation can be examined directly rather than only through a latent representation. The mechanism analysis therefore examines the validation-selected SF--NRM score networks for BTC--RV, ETH--RV, and SOL--RV as representative liquid-asset cases. The diagnostics use the seed-11 fitted model of each selected specification. Because all mechanism diagnostics in this section concern RV, the measure superscript is suppressed throughout Section~\ref{sec:mechanism}.

\subsection{Intraday weight profiles}

\begin{figure}[!htbp]
\centering
\includegraphics[width=0.75\columnwidth]
    {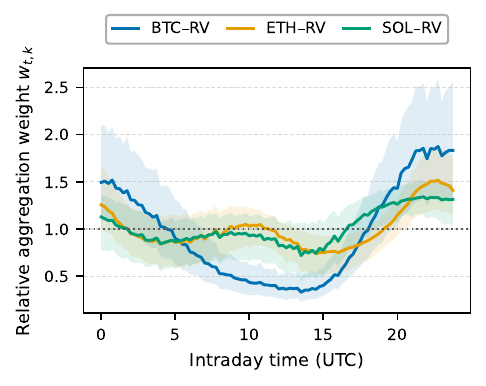}
\Description{Intraday profiles of normalized aggregation weights for BTC, ETH, and SOL realized variance. Lines report test-sample medians, shaded regions report cross-date 10th--90th percentile ranges, and one denotes uniform weighting.}
\caption{Intraday SF--NRM weights for BTC--, ETH--, and SOL--RV. Lines show
medians, bands show 10th--90th percentiles, and one denotes uniform
aggregation.}
\label{fig:btc_eth_sol_rv_intraday_weight_profile}
\end{figure}

The learned aggregation outcome is summarized by Figure~\ref{fig:btc_eth_sol_rv_intraday_weight_profile}. For a complete five-minute day, the displayed normalized weight is $w_{t,k}$. Thus $w_{t,k}=1$ denotes uniform aggregation, values above one indicate relative up-weighting, and values below one indicate relative down-weighting. Five-minute weights are averaged within non-overlapping 15-minute intervals; the lines show cross-date medians and the bands show 10th--90th percentile ranges.

The three assets place relatively less weight on much of the intermediate UTC
day and more weight near the beginning and especially the end of the day.
The common direction is consistent with a shared intraday component, while
differences in level, late-day amplitude, and dispersion indicate that the
learned reallocation is adapted to each asset rather than imposed as a common profile. The late-day pattern may partly reflect the UTC boundary or forecast origin and should not be read as a universal session effect.

\subsection{Local-context and time effects}

The targeted SHAP display \citep{lundberg2017unified} focuses on the
Time and Local groups, which have the largest and most directly
interpretable bar-level effects. The smaller and more measure-dependent
Shape and Global contributions are summarized through the ablation evidence.

\begin{figure}[!htbp]
\centering
\includegraphics[width=0.70\columnwidth]
    {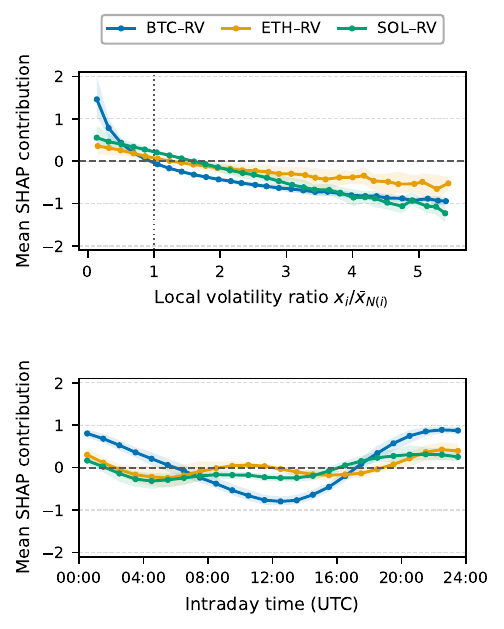}
\Description{Two vertically stacked cross-asset SHAP diagnostics for the
BTC, ETH, and SOL realized-variance score networks. Panel (a) compares the
local-context effect and Panel (b) compares the intraday-time effect. Both
panels use the same vertical scale.}
\caption{SHAP effects of Local and Time inputs on SF--NRM scores. Lines show bin
means and bands show interquartile ranges.}
\label{fig:btc_eth_sol_rv_score_mechanism}
\end{figure}

Figure~\ref{fig:btc_eth_sol_rv_score_mechanism} examines how the Local
and Time inputs affect the fitted pre-normalization scores. It reports
mean SHAP contributions within common bins for representative liquid-asset cases. Panel~(a) displays the local volatility ratio on
its RRM scale, although the network receives its logarithm. Positive
(negative) values indicate inputs that raise (lower) the score relative
to its baseline.

Panel~(a) shows broadly similar nonlinear local-context effects across the
three assets. Locally small realized-variance contributions receive positive
score adjustments, whereas locally elevated contributions generally receive
negative adjustments. The common direction suggests that the reallocation
mechanism is not specific to one cryptocurrency, although differences in
magnitude indicate asset-level heterogeneity.

Panel~(b) identifies a similar but non-identical intraday component in the
three representative assets. The score networks produce more positive SHAP
contributions near parts of the beginning and end of the UTC day and lower
score contributions through much of the intermediate period. Differences in
the timing and amplitude of extrema show that this common structure is
adapted to each asset rather than imposed deterministically.

\section{Conclusion}
\label{sec:conclusion}

This paper studies whether forecast-supervised intraday reallocation can improve volatility prediction while retaining the primitive-level structure of conventional realized measures. The design treats established realized-measure primitives as statistically grounded building blocks and applies neural learning to the aggregation rule rather than replacing those primitives. NRM preserves the chosen intraday primitive, uses mean-normalized positive weights, and exactly recovers the corresponding RRM under uniform allocation. Because the output remains an explicit weighted sum of the original contributions, the daily measure also retains direct primitive-level attribution. NRM therefore provides a structured alternative to both fixed realized-measure aggregation and unrestricted end-to-end neural forecasting.

Controlled simulations show lower reported losses for contemporaneous and next-day latent integrated variance across the evaluated continuous, jump, and conditional event-stress regimes. In cryptocurrency data, paired RRM--NRM comparisons show broad gains under SF, HAR, and MIDAS forecasting specifications, with particularly consistent improvements under QLIKE. Validation-selected NRM specifications attain the lowest mean QLIKE across the six assets among the evaluated approaches. For representative BTC, ETH, and SOL specifications, the fitted score functions combine day-level distributional conditioning, recurring intraday timing, and nonlinear local-context adjustments.

The evidence has several boundaries. NRM is forecasting-specification-specific and should be interpreted as a task-specific representation rather than a reusable universal measure. The cryptocurrency setting provides consistent 24/7 UTC days but does not establish generalization to markets with sessions and overnight gaps. The empirical target is a noisy realized proxy, the simulation regimes are controlled stresses rather than fully calibrated microstructure models, and STLG excludes richer volume and order-book information. Future work can examine transferable aggregation, alternative day boundaries, traditional asset classes, and economic-value applications.


\bibliographystyle{ACM-Reference-Format}
\bibliography{references}

@article{andersen2003modeling,
  author  = {Andersen, T. G. and Bollerslev, T. and Diebold, F. X. and Labys, P.},
  title   = {Modeling and Forecasting Realized Volatility},
  journal = {Econometrica},
  year    = {2003},
  volume  = {71},
  number  = {2},
  pages   = {579--625}
}

@article{andersen2023pockets,
  author  = {Andersen, T. G. and Li, Y. and Todorov, V. and Zhou, B.},
  title   = {Volatility Measurement with Pockets of Extreme Return Persistence},
  journal = {Journal of Econometrics},
  year    = {2023},
  volume  = {237},
  number  = {2},
  pages   = {105439}
}

@article{barndorff2004bipower,
  author  = {Barndorff-Nielsen, O. E. and Shephard, N.},
  title   = {Power and Bipower Variation with Stochastic Volatility and Jumps},
  journal = {Journal of Financial Econometrics},
  year    = {2004},
  volume  = {2},
  number  = {1},
  pages   = {1--37}
}

@article{barndorff2008kernels,
  author  = {Barndorff-Nielsen, O. E. and Hansen, P. R. and Lunde, A. and Shephard, N.},
  title   = {Designing Realized Kernels to Measure the Ex Post Variation of Equity Prices in the Presence of Noise},
  journal = {Econometrica},
  year    = {2008},
  volume  = {76},
  number  = {6},
  pages   = {1481--1536}
}

@inproceedings{chen2016xgboost,
  author    = {Chen, T. and Guestrin, C.},
  title     = {{XGBoost}: A Scalable Tree Boosting System},
  booktitle = {Proceedings of the 22nd ACM SIGKDD International Conference on Knowledge Discovery and Data Mining},
  year      = {2016},
  pages     = {785--794}
}

@article{chan2018stockconnect,
  author  = {Chan, B. S. F. and Cheng, A. C. H. and Ma, A. K. C.},
  title   = {Stock Market Volatility and Trading Volume: A Special Case in Hong Kong with Stock Connect Turnover},
  journal = {Journal of Risk and Financial Management},
  year    = {2018},
  volume  = {11},
  number  = {4},
  pages   = {76}
}

@article{corsi2009har,
  author  = {Corsi, F.},
  title   = {A Simple Approximate Long-Memory Model of Realized Volatility},
  journal = {Journal of Financial Econometrics},
  year    = {2009},
  volume  = {7},
  number  = {2},
  pages   = {174--196}
}

@article{cotturo2026multifactor,
  author  = {Cotturo, P. and Liu, F. and Proner, R.},
  title   = {Multifactor Timing with Deep Learning},
  journal = {Journal of Financial Econometrics},
  year    = {2026},
  volume  = {24},
  number  = {3},
  pages   = {nbag006}
}

@article{diebold1995comparing,
  author  = {Diebold, F. X. and Mariano, R. S.},
  title   = {Comparing Predictive Accuracy},
  journal = {Journal of Business \& Economic Statistics},
  year    = {1995},
  volume  = {13},
  number  = {3},
  pages   = {253--263}
}

@article{garman1980,
  author  = {Garman, M. B. and Klass, M. J.},
  title   = {On the Estimation of Security Price Volatilities from Historical Data},
  journal = {Journal of Business},
  year    = {1980},
  volume  = {53},
  number  = {1},
  pages   = {67--78}
}

@article{ghysels2006midas,
  author  = {Ghysels, E. and Santa-Clara, P. and Valkanov, R.},
  title   = {Predicting Volatility: Getting the Most Out of Return Data Sampled at Different Frequencies},
  journal = {Journal of Econometrics},
  year    = {2006},
  volume  = {131},
  number  = {1--2},
  pages   = {59--95}
}

@article{harvey1997testing,
  author  = {Harvey, D. and Leybourne, S. and Newbold, P.},
  title   = {Testing the Equality of Prediction Mean Squared Errors},
  journal = {International Journal of Forecasting},
  year    = {1997},
  volume  = {13},
  number  = {2},
  pages   = {281--291}
}

@article{heston1993closed,
  author  = {Heston, S. L.},
  title   = {A Closed-Form Solution for Options with Stochastic Volatility with Applications to Bond and Currency Options},
  journal = {The Review of Financial Studies},
  year    = {1993},
  volume  = {6},
  number  = {2},
  pages   = {327--343}
}

@article{li2025wicks,
  author  = {Li, Y. and Nolte, I. and Nolte, S. and Yu, S.},
  title   = {Realized Candlestick Wicks},
  journal = {Journal of Econometrics},
  year    = {2025},
  volume  = {250},
  pages   = {106014}
}

@inproceedings{lundberg2017unified,
  author    = {Lundberg, S. M. and Lee, S.-I.},
  title     = {A Unified Approach to Interpreting Model Predictions},
  booktitle = {Advances in Neural Information Processing Systems},
  year      = {2017},
  volume    = {30}
}

@article{morenopino2024deepvol,
  author  = {Moreno-Pino, F. and Zohren, S.},
  title   = {{DeepVol}: Volatility Forecasting from High-Frequency Data with Dilated Causal Convolutions},
  journal = {Quantitative Finance},
  year    = {2024},
  volume  = {24},
  number  = {8},
  pages   = {1105--1127}
}

@article{patton2011volatility,
  author  = {Patton, A. J.},
  title   = {Volatility Forecast Comparison Using Imperfect Volatility Proxies},
  journal = {Journal of Econometrics},
  year    = {2011},
  volume  = {160},
  number  = {1},
  pages   = {246--256}
}

@inproceedings{kim2024cats,
  author    = {Kim, D. and Park, J. and Lee, J. and Kim, H.},
  title     = {Are Self-Attentions Effective for Time Series Forecasting?},
  booktitle = {Advances in Neural Information Processing Systems},
  year      = {2024},
  volume    = {37}
}

@inproceedings{eldele2024tslanet,
  author    = {Eldele, E. and Ragab, M. and Chen, Z. and Wu, M. and Li, X.},
  title     = {{TSLANet}: Rethinking Transformers for Time Series Representation Learning},
  booktitle = {Proceedings of the 41st International Conference on Machine Learning},
  series    = {Proceedings of Machine Learning Research},
  year      = {2024},
  volume    = {235},
  pages     = {12409--12428}
}

@inproceedings{tan2024language,
  author    = {Tan, M. and Merrill, M. A. and Gupta, V. and Althoff, T. and Hartvigsen, T.},
  title     = {Are Language Models Actually Useful for Time Series Forecasting?},
  booktitle = {Advances in Neural Information Processing Systems},
  year      = {2024},
  volume    = {37}
}

@article{wu2025revisiting,
  author  = {Wu, H.},
  title   = {Revisiting Attention for Multivariate Time Series Forecasting},
  journal = {Proceedings of the AAAI Conference on Artificial Intelligence},
  year    = {2025},
  volume  = {39},
  number  = {20},
  pages   = {21528--21535}
}

@article{martens2007range,
  author  = {Martens, M. and van Dijk, D.},
  title   = {Measuring Volatility with the Realized Range},
  journal = {Journal of Econometrics},
  year    = {2007},
  volume  = {138},
  number  = {1},
  pages   = {181--207}
}

@misc{reisenhofer2022harnet,
  author = {Reisenhofer, R. and Bayer, X. and Hautsch, N.},
  title  = {{HARNet}: A Convolutional Neural Network for Realized Volatility Forecasting},
  year   = {2022},
  note   = {arXiv:2205.07719}
}

@article{rogers1991,
  author  = {Rogers, L. C. G. and Satchell, S. E.},
  title   = {Estimating Variance from High, Low and Closing Prices},
  journal = {Annals of Applied Probability},
  year    = {1991},
  volume  = {1},
  number  = {4},
  pages   = {504--512}
}

@article{zhang2024intraday,
  author  = {Zhang, C. and Zhang, Y. and Cucuringu, M. and Qian, Z.},
  title   = {Volatility Forecasting with Machine Learning and Intraday Commonality},
  journal = {Journal of Financial Econometrics},
  year    = {2024},
  volume  = {22},
  number  = {2},
  pages   = {492--530}
}

@article{fan2017sufficient,
  title   = {Sufficient Forecasting Using Factor Models},
  author  = {Fan, J. and Xue, L. and Yao, J.},
  journal = {Journal of Econometrics},
  volume  = {201},
  number  = {2},
  pages   = {292--306},
  year    = {2017},
  doi     = {10.1016/j.jeconom.2017.08.009}
}

@inproceedings{ilse2018attention,
  title     = {Attention-based Deep Multiple Instance Learning},
  author    = {Ilse, M. and Tomczak, J. and Welling, M.},
  booktitle = {Proceedings of the 35th International Conference on Machine Learning},
  series    = {Proceedings of Machine Learning Research},
  volume    = {80},
  pages     = {2127--2136},
  year      = {2018},
  publisher = {PMLR}
}

@misc{chan2026occupation,
  author = {Chan, B. S. F. and Cheung, K.-F. and Chen, N. and Chan, J. S.-K.},
  title  = {Time-Occupation Information for Realized Volatility Forecasting},
  year   = {2026},
  month  = sep,
  note   = {SSRN working paper}
}

\end{document}